\documentclass{sig-alternate}
\usepackage{url}
\usepackage{multirow}

\usepackage{array}
\usepackage{rotating}
\usepackage{color}
\usepackage{algorithm}
\usepackage{epstopdf}
\usepackage{graphicx}
\usepackage{epstopdf}
\usepackage[noend]{algpseudocode}

\newfont{\mycrnotice}{ptmr8t at 7pt}
\newfont{\myconfname}{ptmri8t at 7pt}
\let\crnotice\mycrnotice%
\let\confname\myconfname%

\permission{Permission to make digital or hard copies of all or part of this work for personal or classroom use is granted without fee
provided that copies are not made or distributed for profit or commercial advantage and that copies bear this notice and the full citation
on the first page. Copyrights for components of this work owned by others than ACM must be honored. Abstracting with credit is permitted. To
copy otherwise, or republish, to post on servers or to redistribute to lists, requires prior specific permission and/or a fee. Request
permissions from Permissions@acm.org.}
\conferenceinfo{SIGKDD'15,}{August 10--13, 2015, Sydney, Australia.}
\copyrightetc{Copyright 2015 ACM \the\acmcopyr}
\crdata{978-1-4503-2956-9/14/08\ ...\$15.00.\\
http://dx.doi.org/10.1145/2623330.2623707}

\begin{document}

\title{On the Formation of Circles in Co-authorship Networks }


\numberofauthors{1} 
\author{\alignauthor Tanmoy Chakraborty$^{1}$\thanks{First two authors have equal contributions.}, Sikhar Patranabis$^{2}$, Pawan
Goyal$^{3}$, Animesh Mukherjee$^{4}$\\
\affaddr{Dept. of Computer Science \& Engg., Indian Institute of Technology, Kharagpur, India -- 721302} \\
\email{\{$^1$its\_tanmoy,$^2$sikharpatranabis,$^3$pawang,$^4$animeshm\}@cse.iitkgp.ernet.in}\\
}

\maketitle
\begin{abstract}

The availability of an overwhelmingly large amount of bibliographic information including citation and co-authorship data makes it
imperative
to have a systematic approach that will enable an author to organize her own personal academic network profitably. An effective method could
be to have one's co-authorship network arranged into a set of ``circles'', which has been a recent practice for organizing relationships
(e.g., friendship) in many online social networks.

In this paper, we propose an unsupervised approach to automatically detect circles in an ego network such that each circle represents a
densely knit community of researchers. Our model is an unsupervised
method which combines a variety of node features and node similarity measures. The model is built from a rich co-authorship network data of
more than 8 hundred thousand authors. In the first level of evaluation, our model achieves 13.33\% improvement in terms of overlapping modularity
compared to
the best among four  state-of-the-art community detection methods. Further,
we conduct a task-based evaluation --  two  basic frameworks for collaboration prediction are considered with the circle information
(obtained from our model)  included in the feature set. Experimental results
show that including
the circle information detected by our model improves the prediction performance by $9.87\%$ and $15.25\%$ on average in
terms of $AUC$ (Area under the
ROC) and $Prec@20$ (Precision at Top 20) respectively compared to the
case, where the circle information is not present. 
\end{abstract}

\section{Introduction}

Now-a-days, public repositories of bibliographic datasets such as DBLP and Google Scholar allow us access
to a stream of scientific articles published by
authors from different domains. An author, we wish to analyze, might be associated with overwhelming volumes of information in terms of her
collaborations and publications, which in turn leads to both {\em information overload} and high {\em computational complexity}. Moreover
from an
author's perspective, it could become painstakingly difficult to keep track of the entire set of academic relationships she has with her
collaborators at
any point of time.\\

\begin{figure}[!h]
\vspace{-0.1cm}
\centering
\includegraphics[scale=0.38]{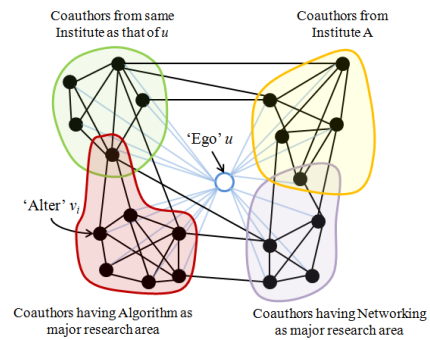}
\caption{(Color online) A hypothetical example showing an ego network of an author $u$ with labeled circles. 
Alters might belong to multiple groups and form overlapping circles.
}
\label{fig:example}
\vspace{-0.2cm}
\end{figure}

\noindent{\bf Present Work: Problem definition.} In this article, we study the problem of automatically
discovering an author's academic circles. In particular, given a single author with her co-authorship network, our goal is to identify her
circles, each of which is a subset of her coauthors. Some examples of real-world circles in an author's co-authorship network are shown in
Figure \ref{fig:example}. The ``owner'' of such
a network (the ``ego'') may wish to form circles based on common bonds and attributes among her coauthors (the ``alters''). An author could
have several reasons behind initiating a new
collaboration. Some common tendencies exhibited by authors include collaborations with the people from her own Institute or with people
sharing the same research interest with her. 
 Therefore, the problem of deciding upon a single dimension to both characterize the
circles and categorize the coauthors appropriately becomes extremely challenging. Moreover, circles are author-specific, as each author
organizes her
personal network of coauthors 
independent of all other authors with
whom she is not connected. This leads to a problem of designing an automatic method that organizes an
author's academic network, more precisely, categorizes her surrounding neighborhoods into meaningful circles.\\

\noindent{\bf Present Work: Motivation of the work.} The problem of detecting ego-centric circles in a co-authorship network is useful in
many
aspects. The collaborators of a particular researcher might have interests aligned with different topics, and the set of collaborators the
researcher
is 
currently working with is a reflection of her current topic of interest. Thus, by understanding circles around her co-authorship network,
she might discover that she might be interested in reading papers about a certain topic that she has not been interested in before. This
result
in turn helps in personalized paper recommendations. On the other hand, if one is interested to start a new collaboration in a particular
field with very famous researcher (usually having less opportunity for new collaboration), a more successful attempt could be to first
establish a collaboration to one of the coauthors of the famous researcher who happens to belong to a circle that is most aligned to the
field of interest. Therefore, the circle information could lead
 to the design of a meaningful collaboration prediction system. Moreover, one can also discover the collaboration pattern of a researcher by
observing the temporal evolution of the ego-centric circles of an author. \\

\noindent{\bf Present Work: An unsupervised approach for circle detection.} In this work, we propose an unsupervised method to learn the
major dimensions of \emph{author profile similarity} that lead to densely linked circles. In practice, since the topological evidence in
such small ego networks is less, the traditional community finding algorithms fail to discover meaningful circles from it
\cite{McAuley,Julian}. Here, we intend the following two conditions to be satisfied during circle detection: first, we expect the circles to
be formed by densely
connected sets of alters. Different circles might overlap, i.e., alters might belong to multiple circles simultaneously. Second, we
expect that the members of the same circle share common properties or traits \cite{mislove2010}.  We model the similarity between
alters as a
function of common profile information. We then propose an {\em unsupervised method} to learn precisely which dimensions of profile
similarity lead to
densely linked circles. In each iteration, our model tries to learn the connectivity between alters from the actual graph and updates the
circle memberships accordingly. Once the optimal condition is encountered, the model outputs the circles. We make our experimental codes
available in the spirit of reproducible research: \url{http://cnerg.org/circle}. \\

\noindent{\bf Summary of the evaluation.} The entire experiment is conducted
on a massive dataset of computer science domain constituting more than 8 hundred thousand authors. Some interesting observations from
the extensive analysis of the detected ego-centric circles are as follows: (i) the highly-cited authors tend to form more number of large
and highly cohesive circles, (ii) the highly-cited authors seem to coauthor with a group of people having a specific research interest in a
particular time period and then leave this group to form another such group of coauthors; (iii) the
highly-cited authors tend to spawn circles that have alters in very similar fields, whereas authors with medium-citations spawn more diverse
circles. To evaluate the quality of the detected circles, we compare our model with four state-of-the-art overlapping community detection
algorithms in terms of standard overlapping modularity
measure and achieve an improvement of $13.33\%$ over the best baseline method. Further, we conduct a task based evaluation where we show
that including the circle information detected by our model in the
feature set improves the performance of the existing collaboration prediction models (liner regression and
supervised random walks) by $9.87\%$ and $15.25\%$ respectively in 
terms of $AUC$ (Area under the ROC curve) and $Prec@20$, compared to the case where the circle information is not present. With 
respect to the best baseline which gives the circle information, our model achieves average improvement of $3.35\%$ and $6.26\%$
respectively in terms of $AUC$ and $Prec@20$. \\

\if{0}
\noindent{\bf Present work. Leveraging ego-centric circles in predicting collaborations.} In the second half of this article, we show how
one can systematically leverage the circle information of each author's ego network in the task of collaboration prediction
\cite{Borner:2005,Eaton}. We choose {\em supervised random walks} as a state-of-the-art collaboration prediction model and explore its
performance with and without the
circle information as one of the features in the feature set. The experiment is repeated for three training sets in three distinct time
points. Experimental results demonstrate that by simultaneously considering node/edge centric features along with the circle information
(whether
two given nodes belong to at least one common circle and, if so, how many common circles do they belong to),
the prediction accuracy achieves an
average $AUC$ (Area under the ROC curve) value $0.832$ and $Prec@20$ value of $0.224$, which are $8.57\%$ and $12.56\%$ higher than the
corresponding values for the baseline model (which precludes circle information).\\

\noindent{\bf Present work: A massive publication dataset.} The entire experiment is conducted on a massive publication dataset of computer
science domain. We crawled all the scientific articles in computer science domain hosted by Microsoft Academic Search. 
Apart from the citation information, each paper is associated with a set of metadata information such as the title of the paper, the
author(s) of the paper, the year and the venue of publication, the related field(s) of research that the paper contributes to, the abstract
and the keyword set of the paper. After a series of preprocessing and filtering steps, we used more than 2 million papers in our
experiments. To the best
of our knowledge, this is the largest bibliographic dataset which contains such a rich metadata information. We have
made the dataset publicly available at \url{http://cnerg.org} (see ``Resources'' tab).

\fi

\vspace{-0.4cm}
\section{Related Work}\label{sec:related_work}
We broadly divide the related work into two subparts: research on the ego structure of a
co-authorship network and research on discovering local circles in an ego network.\\

\if{0}
\subsection{Research on co-authorship networks}
Bibliometric studies have addressed the issue of collaborations profusely descending from global or national views \cite{Braun}
to the individual level \cite{newman04}. Other studies have focused on the influence of the collaboration origins as well as on research
production and its impact. 
There are thus numerous papers that found a positive relationship between the degree of international collaboration and the overall research
impact \cite{tanmoysnam,Katz}, 
whereas others did not find any influence at all \cite{Herbertz}. 
For instance, Stefaniak \cite{Stefaniak} and Lee and Bozeman \cite{Lee} evidenced that ``expensive discipline'' such as Physics, Chemistry
and Biomedicine were 
more collaborative, in spite of the fact that no significant citation difference across disciplines was observed \cite{Vaughan}.
\fi

\noindent{\bf Research exploring ego structure in co-authorship network.} One of the most interesting yet curiously understudied aspects is
the analysis of the structural properties of the ego-alter interactions
in co-authorship networks. Eaton et al. \cite{Eaton} found that the productivity of an author is associated with centrality
degree 
confirming that scientific publishing is related with the extent of collaboration; B\"{o}rner et al. \cite{Borner:2005} presented several
network
measures 
that investigated the changing impact of author-centric networks. Yan and Ding \cite{Yan:2009} analyzed the Library and Information Science
co-authorship network in relation
to the impact of their researchers, finding important correlations. Abbasi et al. 
extensively studied the relationship between scientific impact and co-authorship pattern, discovering significant correlations between
network indicators (density and ego-betweenness) and performance indicators such as g-index \cite{Abbasi:2012} and citation counts
\cite{Abbasi201466}. 
McCarty et
al. 
\cite{McCarty} attempted to predict the h-index evolution through ego networks, observing that this factor increases if one can choose to
coauthor articles with authors already having a high h-index.\\

\noindent\textbf{Research on discovering local circles in ego network}
McAuley and Leskovec \cite{McAuley,Julian} were the first who explored social circles in ego networks. They mapped this problem as a
multi-membership node clustering problem and developed a model for detecting circles that combines network structure as well as user profile
information from Google+, Facebook and Twitter.  
They remarked that these local circles can not
be discovered using traditional community detection algorithms \cite{NewGir04} because of the dearth of information on topological structure
in the ego
network of each author \cite{chakraborty_kdd,Julian}. According to them, under such circumstances topic-modeling techniques
\cite{Airoldi:2008,Ramnath} are the best to
uncover ``mixed memberships'' of nodes to multiple groups. 
\if{0}
The
fundamental differences of our approach compared to theirs are threefold. First, in their paper each circle is explicitly associated
with a profile similarity parameter with respect to which the circle is expected to look similar. In our method, each circle is simply a
collection of similar authors, which allows the circle to be associated with more than one profile similarity  parameter. Moreover, each
circle is associated with a threshold which encompasses all profile parameters. Second, the circles detected by their method are treated as
latent variables and they used coordinate ascent to jointly optimize the set of circles and profile similarity patterns. However, we treat
the authors' membership in various circles as the latent variable and use the simulated annealing concept \cite{Liu20102300} to disturb the
membership and to
arrive at a better configuration. Finally, while they used pseudo-boolean optimization in a
pairwise graphical model, our method uses randomization for perturbing the membership. 
\fi
This, to
the best of our knowledge, is the first attempt to detect local circles (groups of coauthors with similar features) centered around
each ego/author in a {\em co-authorship network} and to use this information further to enhance the performance of existing {\em
collaboration
prediction models}.

\if{0}
The rest of this article is organized as follows. We propose an unsupervised model for the formation of edges within circles in Section
\ref{sec:method} followed by an efficient parameter learning strategy in Section \ref{sec:parameter}. In Section \ref{sec:dataset}, we
describe the dataset used in our experiment.  The set of features for each author that are used in our model to generate the vertex profile
and to measure the author similarity is described in Section \ref{sec:fec_extrac}. A broad empirical analysis and a quantitative
interpretation of
the detected circles are discussed in Section \ref{sec:result} and Section \ref{sec:interpretation} respectively. Finally, in Section
\ref{sec:application} we show how the inclusion of circle information improves the accuracy of a well-known state-of-the-art link
prediction model, supervised random walks. The article is concluded by discussing the pros and cons of our model in Section
\ref{sec:discussion} and the potential future work in Section \ref{sec:conclusion}.
\fi

\section{An Unsupervised Model for discovering ego-centric circles}\label{sec:method}

Our model for detecting ego-centric circles applies to any general ego network, where each node is considered as an ego and the set 
of her one-hop neighbor nodes constitute the set of alters. The ego is said to spawn the ego network, but is not considered
as a part of the network. Our method intends to discover circles in this ego network in an unsupervised fashion, leveraging properties
specific to nodes as well as properties of the network. Our model requires each node to have a profile, which is essentially
the feature vector characterizing the node in a feature space. Two nodes are said to be similar if their feature vectors are similar, as
evaluated
by an appropriate similarity metric. Although exact profile details and the similarity metrics will vary depending on the nature of the
network, some general
assumptions made by our model are as follows:\\
$\bullet$ Alters of the same ego, that have similar profiles should be in the same circles while those with dissimilar profiles should be in
different circles.\\
$\bullet$ Alters that share an edge are more likely to be part of the same circle than disconnected alters.\\
$\bullet$  While it should be possible to label each circle by some common property of its member nodes, a circle may actually have more
than
one label. In our earlier example of a co-authorship network, two or more circles may contain authors from the same field, but may be
different in some other attribute such as the authors from the same Institute as shown in Figure \ref{fig:example}. \\
$\bullet$ Circles may overlap and may even contain other smaller circles.
  
We now describe the algorithm for circle formation  in more details. The input to our algorithm is an ego network $G = <V,E>$.
Each node $v \in V$ has an $N$-dimensional profile vector $F_v$ = \{$f_{1v}$, $f_{2v}$, $f_{3v}$, ..., $f_{Nv}$\},  where $f_{iv}$ 
denotes the value of the $i^{th}$ feature of the node $v$. The ego node $u$, often referred to as the \emph{center} node, is responsible for
spawning the ego network, but does not itself feature as a part of the network. So the ego network of 
$u$ is essentially the subgraph induced by the alters of $u$. Let $D(x,y)$
be the Euclidean distance between the profile vectors of nodes $x$ and $y$ given by Equation \ref{eq:distance}.
\vspace{-0.1cm}
\begin{equation}
\label{eq:distance}
 D(x,y) = D(y,x) = \sqrt{ \sum_{i=1}^{N} (f_{ix}-f_{iy})^2}
\end{equation}

The aim of the method is to identify a set of circles $\hat{C}$= \{$C_1$, $C_2$,.....,$C_K$\}. 
Given a circle $C_j \in {\hat{C}}$ and a
node $y\in V$, we define the distance of $y$ from $C_j$, say $D'(C_j,y)$, as the average distance of $y$ from all other nodes in $C_j$.
Also, the profile similarity measure between a pair of nodes $x$ and $y$, denoted by $Sim(x,y)$ is defined to be the reciprocal of
$D(x,y)$. 
Analogously, the similarity 
between node $y$ and circle $C_j$, denoted by $Sim'(C_j,y)$ is defined to be the reciprocal of $D'(C_j,y)$. 
We shall demonstrate the merit of this profile similarity measure in Sections \ref{sec:parameter} and \ref{sec:fec_extrac}.

Each circle $C_j$ in our model has a similarity threshold parameter $\tau_j$ associated with it such that if node $y \in V$
is in $C_j$ then the following constraint is satisfied:
\begin{equation}
\label{eq:constraint} 
Sim'(C_j,y) \geq \tau_j 
\end{equation}

Based on our assumption that nodes within a common circle at any point of time have a higher probability of forming an edge in the network,
our model predicts the circles estimated at each step to be cliques, and distinct circles not to share any edge at all. 
Given a set of $K$ circles $\hat{C}$= \{$C_1$, $C_2$,.....,$C_K$\},
along with a set of threshold parameters $\hat\tau$ = \{$\tau_1$, $\tau_2$,...,$\tau_K$\} in any iteration of the algorithm, we define a
\emph{closeness}
estimator for a pair of nodes $(x,y) \in V\times V$ in terms of their circle membership, denoted by $\beta(x,y)$.
Let $\beta_1(x,y)$ and $\beta_2(x,y)$ be defined as follows.

\begin{eqnarray}
 \beta_1(x,y) &=& \sum_{C_j: \{x,y\} \subseteq C_j} {(Sim(x,y) - \tau_j + \lambda)}^{-1}\\
\beta_2(x,y) &=& \sum_{C_j: \{x,y\} \nsubseteq C_j} {(Sim(x,y) - \tau_j + \lambda)}^{-1}
\end{eqnarray}

Note that $\{x,y\} \subseteq C_j$ if both $x$ and $y$ are members of the circle $C_j$, while $\{x,y\} \nsubseteq C_j$ 
if $C_j$ does not contain one or both of $x$ and $y$. The constant $\lambda$ is kept large enough to ensure that no term in the summation
is negative and may simply be taken as the maximum of all threshold values, i.e., $max\{\tau_1, \tau_2, ..., \tau_K\}$.
Note that $\beta_1(x,y)$ is high if $x$ and $y$ share common circles with very high thresholds, while $\beta_2(x,y)$ is high if $x$ and $y$
do not share common circles with high thresholds. 

Now, we define the \emph{closeness} estimator $\beta(x,y)$ as follows.
\begin{equation}
\beta(x,y) = \exp\{{[\beta_1(x,y)}]^2 - {[\beta_2(x,y)]}^2\}
\end{equation}

Note that $\beta(x,y)$ is purely a circle-membership based similarity metric
for the pair $(x,y)$, and increases with increase in the \emph{number} and \emph{threshold values} of the \emph{common circles} which $x$
and $y$ are part of. Thus, the 
\emph{closeness} estimator emphasizes not only the common circle memberships of nodes but also the thresholds of the circles they are
part of.

From the \emph{closeness} information so estimated, the probability that the pair $(x,y)$ forms an edge in $G$  is modeled by:
\begin{equation}
\label{eq:prob}
p((x,y)\in E) = \frac{\beta(x,y)}{1+\beta(x,y)}
\end{equation}

Similarly, for the node-pair $(x,y)$ which does not belong to $E$, the probability is estimated as follows:
\begin{equation}
p((x,y)\notin E) = 1- p((x,y)\in E) = \frac{1}{1+\beta(x,y)}
\end{equation}

Quite evidently, $p(x,y)$ increases with increase in $\beta(x,y)$ and is normalized using \emph{add-one smoothing} \cite{Jurafsky:2000}.  
Thus we get a predicted probability of existence for each possible edge in the network given $\hat C$ and $\hat \tau$. The 
rationale underlying the prediction is that the \emph{closeness} of a pair of nodes $(x,y)$ is proportional to the 
similarity of their profiles as well as the number and similarity thresholds of common circles that they are a part of. Now the model must
ensure that this predicted network indeed corresponds to the real network, for which we present 
the following analysis.

Assuming independent generation of each edge in the graph, the joint probability of $G$ and $\hat{C}$ can be written as 
\begin{equation}
\label{eq:probgraph}
P_{\hat \tau}(G; \hat C) = \prod_{(x,y) \in E} p((x,y) \in E) \prod_{(x,y) \notin E}p((x,y) \notin E)
\end{equation}

We define the following notation \ref{eq:short2} for ease of expression:
\begin{equation}
\label{eq:short2}
\phi(x,y) =  \log{(\beta(x,y))} = ({[\beta_1(x,y)}]^2 - {[\beta_2(x,y)]}^2) 
\end{equation}


Taking logarithm of Equation \ref{eq:probgraph}, and using notation \ref{eq:short2} we can express the log likelihood of $G$ given $\hat{C}$
and $\hat{\tau}$ as:
\begin{equation}\label{eq:loglikelihood}
{\scriptsize
\begin{split}
l_{\hat \tau}(G; \hat{C}) &=  \log{(P_{\hat \tau}(G; \hat C))} \\
& = \sum_{(x,y) \in E} \log{(p((x,y) \in E))} + \sum_{(x,y) \notin E} \log{(p((x,y) \notin E))}\\
 & = \sum_{(x,y)\in E}\log{(\beta(x,y))} - \sum_{(x,y) \in V\times V}\log(1+\beta(x,y))\\ 
&  = \sum_{(x,y)\in E}\phi(x,y) - \sum_{(x,y) \in V\times V}\log(1+ \exp\{\phi(x,y)\}) 
\end{split}
}
\end{equation}

The model thus attempts to identify a set of circles $\hat C$ that maximizes $l_{\hat{\tau}}(G; \hat C)$.
In Section \ref{sec:parameter} we describe how this may be achieved by optimizing $\hat \tau$. Also, in Section \ref{sec:fec_extrac}, we
describe how this generic model can be applied to co-authorship networks in particular.

\section{Unsupervised Learning of Model Parameters}\label{sec:parameter}
In this section, we describe the method used to find the set of circles $\hat{C}$ by maximizing the log likelihood in Equation 
\ref{eq:loglikelihood}. Algorithm \ref{alg:circleupdate} summarizes the steps of a single iteration of the algorithm.

Initially, each node is in a different circle with a very high threshold value. At each iteration $t$,
for each node $y \in V$ we alter the circle membership of $y$ by randomly
adding it to some circles it previously did not belong to and deleting it from some circles it belonged to. This is similar to the concept
of perturbing the solution in simulated annealing \cite{Liu20102300}. 
The circle thresholds are then updated accordingly such that the constraint in Equation \ref{eq:constraint} is not violated.

The general idea is that larger the number of circles a node $y$ is already part of after time step $t$, lesser is the extent to which the
circle membership of $y$ is disturbed in time step $t+1$.

We denote by $\hat{C_{t}}$ the set of circles and by $\hat{\tau _t}$ the corresponding set of thresholds
after time step $t$, where $\hat{C_{t}}$ = \{$C_1(t)$, $C_2(t)$,...,$C_K(t)$\} and
$\hat {\tau_{t}}$ = \{$\tau_1(t)$, $\tau_2(t)$,...,$\tau_K(t)$\}.
Also, let the log likelihood of $G$ given $\hat {C_{t}}$ and $\hat {\tau_t}$ be $l_{\hat{\tau}}(G; \hat{C_{t}})$. 
The following are the main steps of the algorithm to update the circle in time step $t+1$:\\

\textbf{Step 1:} For each node $y\in {V}$, we capture the circle membership of $y$ at time $t$ by defining two sets ${S1}_{y,t}$ and
${S2}_{y,t}$:

\begin{eqnarray}
\vspace{-0.2cm}
{S1}_{y,t} &=& \{C_j(t) | C_j(t) \in \hat{C}_{t} \wedge y \in C_j(t)\}\\
{S2}_{y,t} &=& \{C_j(t) | C_j(t) \in \hat{C}_{t} \wedge y \notin C_j(t)\}
\end{eqnarray}

\textbf{Step 2:} Now we intend to compute the number of circles to add $y$ to and to remove $y$ from, given by the two variables -
${AddCircle(y,t+1)}$ and ${RemoveCircle(y,t+1)}$:
\vspace{-0.2cm}
\begin{eqnarray}
{AddCircle(y,t+1)} &=&  \left \lceil{ \frac{K1 + |{S1}_{y,t}|}{|{S1}_{y,t}|} }\right \rceil \\
{RemoveCircle(y,t+1)} &=&  \left \lceil{ \frac{K2 + |{S1}_{y,t}|}{|{S1}_{y,t}|} }\right \rceil  
\end{eqnarray}

Here, $K1$ is a randomly chosen integer with
$1\leq K1 < |{S2}_{y,t}| $, such that the value of ${AddCircle(y,t+1)}$ is less than or equal to $|{S2}_{y,t}|$, i.e., the
number of circles that $y$ is currently not part of. Similarly, $K2$ is a randomly chosen integer with $1\leq K2 < |{S1}_{y,t}|$ such that
the value of ${RemoveCircle(y,t+1)}$ is less than or equal to $|{S1}_{y,t}|$, i.e., the
number of circles that $y$ is currently part of. Note that both ${AddCircle(y,t+1)}$ and ${RemoveCircle(y,t+1)}$ are low for high values of
$|{S1}_{y,t}|$. This ensures that the more the number of circles $y$ is currently part of, lesser is the disturbance to the circle
membership of $y$ (and vice versa).\\

\textbf{Step 3:} Add $y$ to $AddCircle(y,t+1)$ many randomly chosen circles from ${S2}_{y,t}$ 
and remove $y$ from $RemoveCircle(y,t+1)$ many randomly chosen circles from ${S1}_{y,t}$. The corresponding circles
are updated accordingly.\\

\textbf{Step 4:} Once Steps 1, 2 and 3 are over for each node, we have the set $\hat{C}_{t+1}$ containing
the augmented circles. Next, we update the corresponding thresholds by setting $\tau_j(t+1)$ corresponding to the circle $C_j(t+1)$ to the
minimum value such that for each node $y\in C_j(t+1)$  
the constraint in Equation \ref{eq:constraint} is not violated. Thus the updated $\tau_j(t+1)$ for $C_j(t+1)$ is given by:
\vspace{-0.2cm}
\begin{equation}
\tau_j(t+1) = min \{Sim'(C_j(t+1),y) | y\in C_j(t+1)\} 
\end{equation}

\textbf{Step 5:} If the threshold $\tau_j(t+1)$ for $C_j(t+1)$ falls below a constant lower limit $\tau_L$, we discard $C_j(t+1)$. 
The value of $\tau_L$ is empirically determined. In our experiments, we tested over a wide range of $\tau_L$ and set it to $0.2$ for best
results (see Figure \ref{fig:tau}).\\

\textbf{Step 6:} We then compute the log likelihood $l_{\hat\tau_{t+1}}(G; \hat C_{t+1})$  
using Equation \ref{eq:loglikelihood}. 
If $l_{\hat \tau_{t+1}}(G; \hat C_{t+1}) > l_{\hat \tau_{t}}(G; \hat C_t)$, then retain newly computed sets $\hat C_{t+1}$ and $\hat
\tau_{t+1}$; else set $\hat C_{t+1} = \hat C_{t}$ and $\hat \tau_{t+1} = \hat \tau_{t}$.  

The process continues till we reach a maxima and the log likelihood does not increase any further for 
sufficiently many iterations. We then report the set of circles so obtained as the optimal set of circles. 
Note that the maximum number of circles after any iteration of the algorithm is $|V|$ and the maximum number of nodes in
any circle is also $|V|$. So the running time of each iteration of the algorithm is $O({|V|} + {|C_{t}|}) = O({|V|})$. Also, any change to
the set of circles
is accepted only if the overall likelihood increases and so the method converges to a local maxima after a finite number of steps. For
practical applications, the method is assumed to reach a local maxima if the likelihood function does not increase for $|V|$
iterations. 

\begin{algorithm}
\caption{Iteration for Updating Circles}
\label{alg:circleupdate}
{\scriptsize
\begin{algorithmic}[1]

\Procedure{CircleUpdate}{$t$, $\hat C_{t}$, $\hat \tau_t$, $l_{\hat \tau_{t}}(G; \hat C_{t})$}
\State $\hat C_{t+1} \gets \hat C_{t}$
\State $\hat \tau_{t+1} \gets \hat \tau_{t}$
\ForAll {$y\in {V}$}
\State ${S1}_{y,t} = \{C_j(t) | C_j(t) \in \hat{C}_{t} \wedge y \in C_j(t)\}$
\State ${S2}_{y,t} = \{C_j(t) | C_j(t) \in \hat{C}_{t} \wedge y \notin C_j(t)\}$
\State $K1 \gets random(1,|{S2}_{y,t}|) $
\State $K2 \gets random(1,|{S1}_{y,t}|) $
\State $AddCircle(y,t+1) \gets  \left \lceil{ \frac{K1 + |{S1}_{y,t}|}{|{S1}_{y,t}|} }\right \rceil$
\State $RemoveCircle(y,t+1) \gets \left \lceil{ \frac{K2 + |{S1}_{y,t}|}{|{S1}_{y,t}|} }\right \rceil$
\State \text{Randomly choose} $C_{AC},C_{RC}$:
\State $C_{AC} \subseteq {S2}_{y,t},|C_{AC}| = AddCircle(y,t+1) $
\ForAll{$C_j(t) \in C_{AC}$}
\State $C_j(t+1) \gets C_j(t+1)\cup \{y\}$
\EndFor
\State $C_{RC} \subseteq {S1}_{y,t},|C_{RC}| = RemoveCircle(y,t+1) $
\ForAll{$C_j(t) \in C_{RC}$}
\State $C_j(t+1) \gets C_j(t+1)\smallsetminus \{y\}$
\EndFor
\EndFor
\ForAll{$C_j(t+1) \in \hat C_{t+1}$}
\State $\tau_j(t+1) \gets min \{Sim'(C_j(t+1),y) | y\in C_j(t+1)\}$
\If {$\tau_j(t+1) < \tau_L $ }
\State $\hat C_{t+1} \gets \hat C_{t+1} \smallsetminus \{C_j(t+1)\}$
\State $\hat \tau_{t+1} \gets \hat \tau_{t+1} \smallsetminus \{\tau_j(t+1)\}$
\EndIf
\EndFor.
\State \text{Compute} $l_{\hat \tau_{t+1}}(G; \hat C_{t+1})$ [Eq. \ref{eq:loglikelihood}] 
\If{$l_{\hat \tau_{t+1}}(G; \hat C_{t+1}) > l_{\hat \tau_{t}}(G; \hat C_{t})$}
\State \text{Return} \{$\hat C_{t+1}, \hat \tau_{t+1}\}$
\Else
\State \text{Return} \{$\hat C_{t}, \hat \tau_{t}\}$
\EndIf

\EndProcedure
\end{algorithmic}}
\end{algorithm}
\vspace{-0.2cm}

\section{A Large publication dataset}\label{sec:dataset}
We have crawled one of the largest publicly available data-\\sets from Microsoft Academic Search (MAS) which houses over 4.1 million
publications and 2.7 million authors. We collected all the papers
specifically published in the computer science domain and indexed by MAS. The crawled dataset contains more than 2 million distinct papers
by more than 8 hundred thousand authors, which are further distributed over 24 fields
of computer science domain. The co-authorship network constructed from this dataset has authors as nodes and edges between authors who have
written at least one paper together.


\noindent{\bf Ego network:} The next step is the construction of
ego networks from the co-authorship network. 
We consider the ego networks corresponding to each node (author) present in our dataset, thus obtaining 821,633  ego networks. An
illustrative example of an ego network is shown in Figure \ref{fig:example}. However, in this experiment we consider only the induced
subgraph of the alters for an ego and exclude the ego and its attached edges from the ego network, as mentioned earlier.

\section{Feature Extraction}\label{sec:fec_extrac}

Profile information of each author node in the ego network is
represented as a feature vector consisting of a set of features. 
These features can be divided into two broad categories -- \emph{general} and \emph{ego-centric} features. Having these two 
separate categories, the  feature set emphasizes the fact that members of common circles should not only have high feature similarity
with each other but also
share similar relationships with the ego.

Given an author $x$ with all her publications, and the set of fields of research $F = \{r_1, r_2, ....., r_{24}\}$\footnote{Note that there
are 24 research fields present in our dataset.}, we define the 
\emph{versatility vector} $\hat{V}(x)$ of an author $x$ as  $\{r_{i,x}
; r_i \in F\}$ such that
$r_{i,x}$ is the fraction of publications of $x$ in field $r_i$. Also, given a set of decades $DEC$ = \{1960-1970, 1971-1980, 1981-1990,
1991-2000, 2001-2009\}, we define the \emph{persistence vector}
$\hat{D}(x)$ for $x$ as $\{d_{j,x}; 1\leq j \leq 5\}$, where $d_{j,x}$ denotes the number of papers 
published by $x$ in decade $DEC(j)$. We also define the major field of
work $R(x)$ for $x$, where she has maximum number
of publications.

The \emph{general} features capture independent characteristics of each author in the ego network and are listed below:\\
$\bullet$ The normalized number of citations the author has received (size 1)\\
$\bullet$ The normalized number of citations \emph{per paper} that the author has received (size 1)\\
$\bullet$ The normalized h-index of the author (size 1)\\
$\bullet$ The normalized number of coauthors of the author (size 1)\\
$\bullet$ The \emph{versatility vector} of the author (size 24)\\
$\bullet$ The normalized number of papers written by the author (size 1)\\
$\bullet$ The \emph{persistence vector} of the author (size 5)\\
$\bullet$ The major field of the author (size 1)\\

On the other hand, the \emph{ego-centric features} capture the relationship of an alter with its ego.
Such features include:\\
$\bullet$ The fraction of papers coauthored by the alter with the ego in each of the five decades (size 5)\\
$\bullet$ The fraction of papers coauthored by the alter with the ego in each of the 24 fields (size 24)\\
$\bullet$ The normalized number of common coauthors that the alter has with the ego (size 1)\\
$\bullet$ The fraction of papers authored by the alter in the major field of the ego (size 1)\\
$\bullet$ The fraction of papers authored by the ego in the major field of the alter (size 1)\\

Thus the dimension of the feature vector containing all the above listed features is $67$. Using the profile information for each node, our
model computes the probability of edge existence between each pair of nodes $(x,y)$,
given by $p(x,y)$ as described in Equation \ref{eq:prob}. We calculate this probability from the \emph{extent of similarity} of node-pair
$(x,y)$, i.e., $Sim(x,y)$.
In order to verify that the \emph{node similarity} indeed helps identify edges between similar authors with high probability
of collaboration, we perform two small experiments. We first check the conditional probability that
given a node-pair $(x,y)$ with similarity $Sim(x,y)=W_{xy}$ in
$G(V,E)$, the node-pair indeed materializes as an edge in the real
network. 

\if{0}
Formally, the 
conditional probability $P((x,y)\in E|W_{xy})$ is computed as:
\begin{equation}
\begin{split}
\label{eq:condprob}
P((x,y)\in E|W_{xy}) =\frac{|{(x',y') \in E | Sim(x',y') = W_{xy}}|}{|{(x',y') \in V \times V|Sim(x',y')=W_{xy}}|}
\end{split}
\end{equation}
\fi

\begin{figure}[!h]
\centering
\includegraphics[width=\columnwidth]{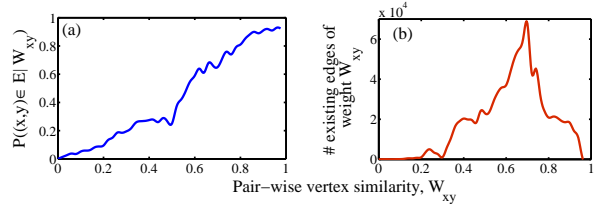}
\caption{(a) Conditional probability of edge existence between authors with a given similarity $W_{xy}$ between their profiles; (b) actual
number of edges with a given edge
weight in the network.}
\label{fig:edge_weight}
\end{figure}

Plot shown in Figure \ref{fig:edge_weight}(a) confirms that our
similarity measure is indeed proportional to the conditional
probability of edge existence. We also observe the number of edges in
a network having a particular edge weight $W_{xy}$ in Figure
\ref{fig:edge_weight}(b). Most of the edges are in the range $0.55 -
0.75$, indicating that this is the most common profile similarity
range among pair-wise authors. Very few edges exist in
the range $0-0.3$, which indicates that collaboration between
authors with very dissimilar profiles is quite rare. This value also seems quite low for the range $0.9-1.0$, which might be due
to the fact that it is extremely rare to have authors with nearly
similar profiles. 

%

\begin{figure}[!h]
\centering
\includegraphics[width=\columnwidth]{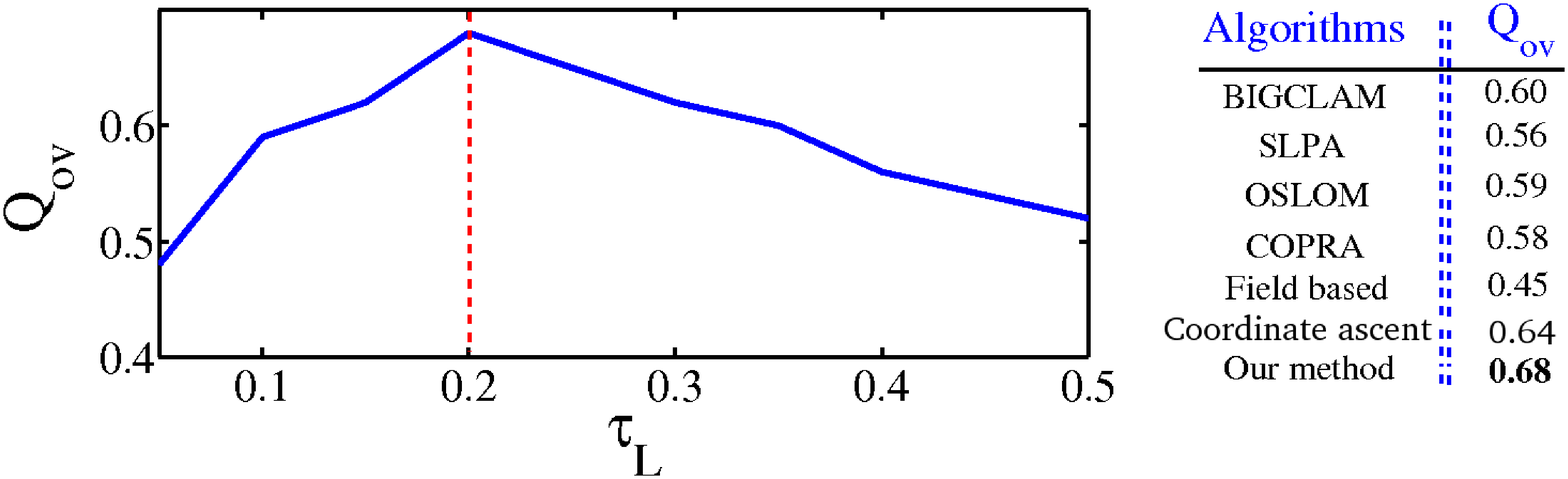}
\caption{(Left) Change in overlapping modularity $Q_{ov}$ with the increase in $\tau_L$; (Right) comparison of the baseline algorithms with
our
method.}\label{fig:tau}
\end{figure}

\begin{figure*}[!ht]
\centering
\includegraphics[scale=0.35]{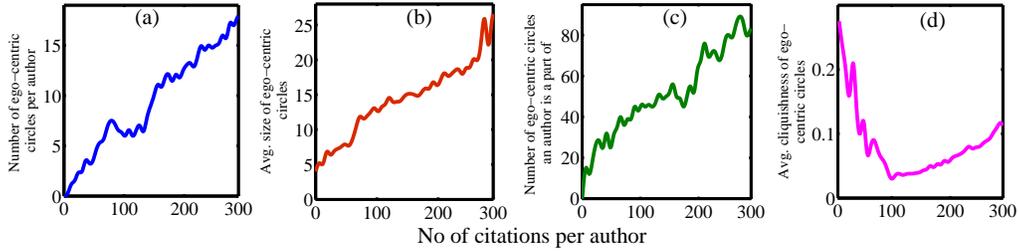}
\caption{(Color online) Author-specific characterization of detected ego-centric circles. The following are plotted against the number of
citations per author :
(a) the number of circles centered around an author, (b) the average size of the circles centered around an author, (c) the number of
ego-centric circles an author is a part of and (d) the average cliquishness of the circles centered around an author.}
\label{fig:author_specific}
\vspace{-0.3cm}
\end{figure*}

\section{Evaluation of detected circles}

In this section, we intend to evaluate the quality of the circles detected by our proposed methodology. Evaluation is especially important
to judge the quality of the detected circles.   
We compare the circles detected by our model with that obtained from four other recent overlapping community detection algorithms, namely
BIGCLAM
\cite{Leskovec}, SLPA
\cite{Xie}, OSLOM
\cite{oslom} and COPRA
\cite{Gregory1}. We also detect the circles using the coordinate ascent method (CA) \cite{McAuley,Julian}. Since we intend to show that
research field of the authors is not the proper information for creating the circles, we also compare our output with the circles obtained
simply from research fields. For comparison, we use {\em overlapping modularity} $Q_{ov}$ \cite{Shen} which is probably the most widely used
measure for evaluating the goodness of a community structure without a ground-truth. 


First, to show the change in $Q_{ov}$ with respect to the threshold $\tau_L$ as described in Section \ref{sec:parameter}, we plot this
quality function in
Figure \ref{fig:tau} by varying $\tau_L$ from 0.05-0.5. We observe that $Q_{ov}$ reaches maximum at $\tau_L=0.2$. Then for each competing
algorithm, we measure the value of $Q_{ov}$ for each ego and take an average over all the egos present in our dataset. The table adjacent to
Figure
\ref{fig:tau} shows that our method outperforms the traditional topology based community finding algorithms in detecting meaningful circles.
Our
method
achieves $Q_{ov}$ of 0.68 which is 6.25\% higher than coordinate ascent method, 13.33\% higher than BIGCLAM, 15.25\% higher than OSLOM,
17.24\% higher than COPRA, and 21.42\% higher than
SLPA. We notice that research field based circles are the worst among the detected circles (see Section \ref{sec:interpretation} for more
discussion).

 \begin{figure*}[!ht]
\centering
\includegraphics[scale=0.35]{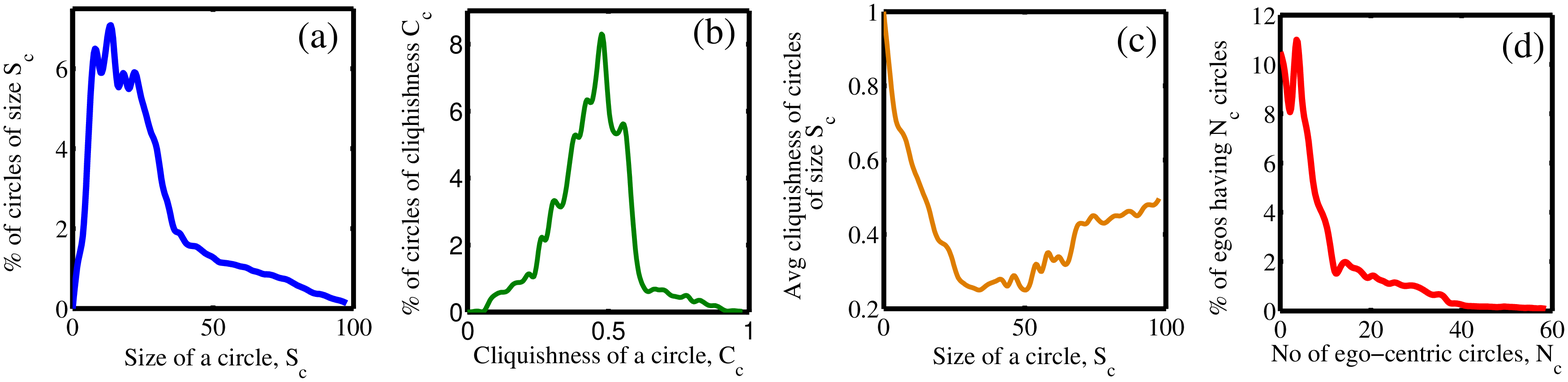}
\caption{(Color online) Circle-specific characteristics of detected ego-centric circles: (a) distribution of the size of the ego-centric
circles, (b) distribution of the cliquishness of the ego-centric circles, (c) average cliquishness of the circles having a particular size,
and (d) percentage of egos surrounded by a particular number of ego-centric circles.}
\label{fig:circle_specific}
\vspace{-0.4cm}
\end{figure*}

\vspace{-3mm}
\section{Analysis of Ego-Centric Circles}\label{sec:result}
In this section, we intend to characterize the ego-centric circles obtained from our unsupervised model. In particular, we study the
properties of the ego-centric circles at two levels of granularity: author-specific analysis and circle-specific analysis.

\subsection{Author-specific Analysis}
Here we study how the circles in the ego networks of the highly-cited authors differ from those of the low-cited authors. Figure
\ref{fig:author_specific}(a) shows the number of ego-centric circles appearing in the ego network of each author. Note that in all of
the experiments, we categorize authors
into three groups: authors receiving more than 100 citations as {\em highly-cited authors} (proportion: 5.21\%), authors receiving
citations between 30-100 as {\em medium-cited authors} (proportion: 28.75\%), and authors receiving less than 30 citations as {\em low-cited
authors} (proportion: 66.04\%). We notice a rise in the number of circles with the increase of citations. The possible reason
could be that since the authors accumulating high citations tend to have high number of collaborators, the number of alters in their ego
networks is also high, and thus more number of ego-centric circles are detected for the highly-cited authors.

In Figure \ref{fig:author_specific}(b), we plot the average size (measured in terms of the number of nodes) of the ego-centric circles  for
the authors in different citation range. 
Once again, the average size of the circles increases with the increase of citations per author. 
It essentially indicates that for the highly-cited egos, the alters are not only high in number but also form large cohesive groups. 

Since each author is also an alter in her neighbors' ego networks, 
she might be a part of multiple such ego-centric circles.
Figure \ref{fig:author_specific}(c) shows the number of such ego-centric circles to which an author belongs to. 
This plot highly correlates with Figure \ref{fig:author_specific}(a), and shows that since highly-cited authors have more number of alters
in their ego networks, 
each of them also belongs to multiple local circles of her neighbors' ego networks.
 
Further, we measure the {\em degree of cliquishness} (the ratio of the number of existing edges in the circle and the maximum number of
possible edges in the circle) of each ego-centric circle. For each ego, we
measure the average
cliquishness of her surrounding circles in the ego network. Figure \ref{fig:author_specific}(d) shows that the average value of
cliquishness initially decreases with the increase of the number of citations per author, then it starts increasing. The reason could be
that
since both the number and the size of the ego-centric circles for low-cited authors are less, the maximum number of possible edges within a
circle is also less, which in turn acts as the reason of high cliquishness for low-cited
authors. In the middle
citation zone, both the number and the average size of the circles are moderate. However, the number of edges that materialize within these
circles is less as compared to the maximum number of possible edges, thus accounting for the sparseness of these circles. Therefore, the
value of cliquishness of circles spawned
by authors
in the middle range of citations is comparatively low. 
However, the value of cliquishness starts increasing for the authors having citations more than 100. 
This signifies that for the highly-cited authors, despite the apparently large size of circles, the probability that an edge 
actually materializes in the real network tends to increase. This explains the formation of dense ego-centric cliques surrounding the ego in
the high-citation range.

\subsection{Circle-specific Analysis}
Now we look into some of the characteristic features specific to an ego-centric circle. 
In Figure \ref{fig:circle_specific}(a), we plot the percentage of ego-centric circles having a particular size $S_c$. 
It follows a Gaussian distribution at the beginning along with a heavy tail at the end. 
We observe that around 65.26\% circles have sizes ranging between 4-30. However, the flat tail at the end shows that more than 15\% circles
have size greater than 50. Figure \ref{fig:circle_specific}(b) shows the distribution of the cliquishness ($C_c$) of the ego-centric
circles. Surprisingly, it again follows a Gaussian distribution with mean $\sim 0.44$ and variance $\sim 0.02$. We notice that around
59.28\% circles
have
cliquishness values ranging between $[0.4,0.6]$ which is quite high. Further inspection reveals that low-degree egos are surrounded by
small-size circles and therefore their cliquishness value is quite high. 
To get a clear idea of the relation between the size and the cliquishness of the ego-centric circles, 
we plot in Figure \ref {fig:circle_specific}(c) the average cliquishness of the circles having a specific size. 
The value of cliquishness $C_c$ gradually decreases with the increase of the size $S_c$ till $S_c$=40, which is followed by a sharp
increase. As mentioned earlier, the increase of cliquishness at the end once again emphasizes that the large-size circles centered around
the highly-cited authors are relatively dense. Finally, we  plot the percentage of egos surrounded by a
specific number of circles in Figure \ref{fig:circle_specific}(d). As expected, we observe that
the plot has a declining trend from the very beginning, which once again highlights our previous observation that
most of the low-cited authors have a low degree in the co-authorship network,
and spawn only a few ego-centric circles.
Since the co-authorship network is mostly dominated by low-degree authors, most of the egos are fringed by a small number of local
circles.

 \begin{figure}[!h]
\centering
\includegraphics[width=\columnwidth]{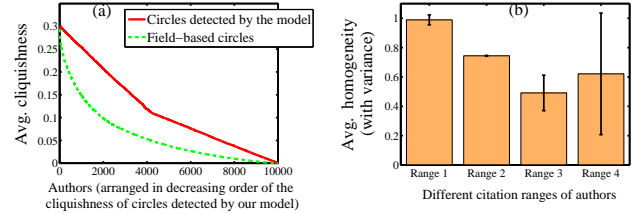}
\caption{(Color online) (a) Comparison of cliquishness of area-based circles and the circles detected by our model, (b) average homogeneity
of
the ego-centric
circles detected by our model for the authors categorized into four zones as per number of citations: Range 1 (>200), Range 2 (>100 \&
<=200), Range 3 (>30 \& <=100)
and Range 4 (<=30).}\label{fig:purity}
\vspace{-0.3cm}
\end{figure}

\section{Interpretation of ego-centric circles}\label{sec:interpretation}

In co-authorship network, most intuitive ground-truth
communities are often assumed to be different areas of research \cite{tanmoysnam} in a particular domain. Therefore, one can
interpret each ego-centric circle as a group of coauthors working in a specific research area. Since
we know the major research area of each author in the dataset, for each ego we further group its coauthors based on only their major
research area such
that each circle corresponds to an area and constitutes coauthors working on this area. Then we measure the average cliquishness of the
research field based circles for each author vis-a-vis that of the ego-centric circles detected by our model. Essentially, we intend to
cross-validate our hypothesis that considering a single dimensional feature vector of an author such as the field information is not an
appropriate way
of encircling alters; rather each circle might represent individual dimension of cohesiveness among its constituent nodes as shown in Figure
\ref{fig:example}. In Figure \ref{fig:purity}(a), we plot the average cliquishness of field-based circles vis-a-vis that of the circles
detected by the model surrounding each
author. As expected, the cliquishness of the detected circles is significantly higher than that of the field-based circles. Therefore, we
conclude that the field-based circles might not appropriately group highly cohesive nodes, rather the circles detected by our model seem to
be more representative and meaningful.

\begin{figure}[!t]
\centering
\includegraphics[scale=0.22]{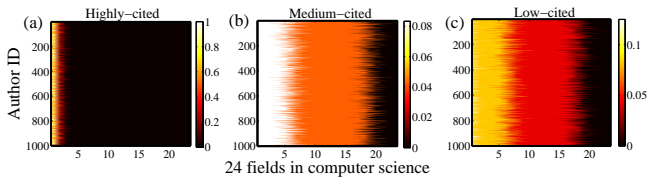}
\caption{(Color online) Heat maps representing the fraction of ego-centric circles marked by each of the 24 fields for highly-cited,
medium-cited and low-cited authors. For each author, the elements in the horizontal axis are sorted in descending order.}
\label{fig:heatmap}
\vspace{-0.2cm}
\end{figure}

\begin{figure*}[!ht]
\centering
\includegraphics[scale=0.32]{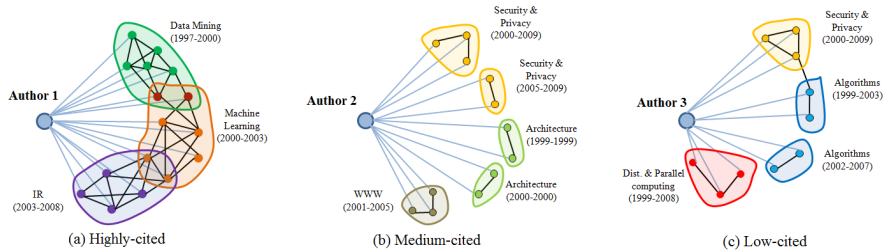}
\caption{(Color online) Examples of ego-networks from three citation zones. Individual nodes have different colors corresponding
to different areas of research. If the color of all the nodes in a circle is same as the color of
the circle, the value of homogeneity is 1. Time period $(t_i-t_j)$ associated with each circle indicates that the ego has written a
paper first (last) time with anyone of its constituent coauthors at year $t_i$ ($t_j$).}\label{fig:sample}
\vspace{-0.4cm}
\end{figure*}

 We further mark each of the detected circles by that field which is also the major research
area for most of its constituent coauthors. Then for each ego, we measure the fraction of circles belonging to each of the 24 fields.
Therefore, each ego/author can now be represented by a vector of size 24 whose $i^{th}$ entry represents the fraction of ego-centric circles
marked by field $i$. Figure \ref{fig:heatmap} shows three heat maps corresponding to highly-cited,
medium-cited and low-cited authors.  For the sake of brevity, we only plot values for 1000 authors from each citation range although the
results are similar for other authors. We observe that for highly-cited authors, ego-centric circles are mostly marked
by few fields, which indicates that the highly-cited authors tend to collaborate with people from similar research area. If this is
true, then the immediate question would be why the coauthors having same research interest are encircled into different groups by our model.
Further
inspection reveals that along with the field, each group also represents the time of collaboration.  For instance, the ego network of
Author 1\footnote{The names of the authors are anonymized in order to maintain privacy.} (one of the highly-cited authors) is shown in
Figure
\ref{fig:sample}(a). One group of her ego network encircles authors in Data Mining  who have coauthored with her during 1997--2000.
Another such group constitutes authors from Machine Learning, who have collaborated with her during 2000--2003.   Therefore, the field
of research and the time of collaboration act as two major dimensions in this case.

Next for the medium-cited authors, the heat map in Figure \ref{fig:heatmap}(b) shows that the distribution of circles into different fields
seems to be much more uniform as compared to Figure \ref{fig:heatmap}(a). The example shown in Figure \ref{fig:sample}(b) also corroborates
with the hypothesis
that with the decrease of citations, the ego-centric circles tend to become even more complicated to be interpreted distinctly. From Figure
\ref{fig:sample}(b), we notice that the time duration of collaboration corresponding to the circles are overlapping and, therefore, it is
very hard
to distinguish these circles. The
result is even more cluttered for low-cited authors as shown in Figures \ref{fig:heatmap}(c) and \ref{fig:sample}(c). These results thus
lead to a general conclusion that the highly-cited authors seem to coauthor with a group of people having a specific research interest in a
particular time period and then move to another such group of coauthors; whereas this tendency is not so prominent for
medium- and low-cited authors. However, we
posit that there might be other dimensions (such as the name of the Institute where the author belongs to) that might
help us interpret these circles more clearly. \\

\noindent{\bf Homogeneity of ego-centric circles.} 
We define a field-based homogeneity for ego-centric circles to verify if, in most cases, authors from the same field tend to
form communities and whether the circles spawned by our unsupervised approach are able to capture this tendency. Given a circle $C$, we
define $F_{C,i}$ to be the fraction of authors in $C$ with major field $f_i$. One can easily 
infer that a uniform distribution $F_{C,i}$ implies that the circle is homogeneous with respect to the field of work
while a skewed distribution (with majority of authors in one or two fields) characterizes a more field-specific circle.
In particular, we define the homogeneity coefficient $H(C)$ for circle $C$  in terms of the entropy of the circle with respect to the
distribution across different fields as in Equation \ref{entro}. Greater the entropy, lesser is the homogeneity and vice versa. 

{\scriptsize
\begin{equation}\label{entro}
  H(C) = \frac{1}{1 -\sum_{i=1}^{24}{F_{C,i}\log(F_{C,i})}}
\end{equation}}


Figure \ref{fig:purity}(b) captures the average homogeneity of circles in the ego-network of authors in four different citation ranges. We
note that
the homogeneity is highest for the authors with very high citation ranges $(>200)$ and has low variance, indicating that
highly-cited authors tend to spawn circles that have alters in very similar fields, whereas authors with medium citations $(30-100)$
spawn more diverse circles. The authors with low citations $(\leq 30)$ exhibit higher degree of homogeneity than those in the medium range,
but
this may be attributed to the fact that they spawn very small-sized circles.

\section{Task based evaluation}\label{sec:application}

We further evaluate the quality of the circles through a task based evaluation framework -- the task of collaboration prediction. We choose
two
supervised learning models: linear regression (LR) \cite{Backstrom} and supervised random walks (SRW) \cite{Backstrom}.
Then we demonstrate that inclusion of the ego-centric circles detected by our model as a feature in the feature set would
eventually enhance the performance of this model with respect to the one in which the circle information is missing. 

\subsection{Problem Definition} For our problem, we assume a temporal graph $G_T=(V_T,E_T)$ where $V_T$ represents a set of nodes such that
each node
$u_t \in V_T$ is associated with a time stamp $t$ indicating its first appearance in $G_T$, and each edge
$e_{t_i,t_j}=(u_{t_i},v_{t_j})$ connects two nodes $u_{t_i}$ and $v_{t_j}$ (such that $u_{t_i}$,$v_{t_j} \in V_T$ and $t_i<=t_j$). Each
node $u_t$ is also associated with a feature vector $f_{u_t}$ at time stamp $t$, whose entires might change over time. Now, given a
longitudinal snapshot of the graph $G_T$ from the beginning till time $T^{'}$, say $G_{T^{'}}=(V_{T^{'}},E_{T^{'}})$, the collaboration
prediction problem aims at predicting the collaborations which are going to appear among the vertices in $V_{T^{'}}$ within $\Delta t$ time
period after $T^{'}$. 

This task is very challenging due to extreme sparsity of real
networks where each node is connected to only a very small fraction of all other nodes in the network (the presence of high proportion of
negative evidences in the dataset).

\if{0}
\subsection{Supervised Random Walks (SRW)} 
Backstrom and Leskovec \cite{Backstrom} proposed a link prediction method based on {\em
supervised
random walks} which combines the information of the network structure with node and edge attributes to predict links efficiently. These
node and edge features are used to learn edge strengths (i.e., random walk transition probabilities) such that a random walk on the network
is more likely to visit ``positive'' than ``negative'' nodes. Here, positive nodes are the ones to which new edges will be created in the
future, and the negative are the remaining nodes. The authors summarized a supervised learning task provided a source node $s$ and training
examples as to which nodes $s$ is connected. Next, they formulated the problem of studying the function that
assigns a strength (i.e., random walk transition probability) to each edge such that the random walk scores in the nodes to which
$s$ creates new links are higher than nodes to which $s$ does not create links. Two aspects behind this model are discussed as follows:

\begin{itemize}
\item {\em Node ranking aspect.} Designing an approach that will assign high scores to nodes which $s$ creates links to than to those that
$s$
does not link to. This method mainly takes advantage of the structure of the network.

 \item {\em Classification aspect.} Exploiting a classifier that predicts pairs of nodes that $s$ is going to create links with and denote
them as
positive training examples; the other nodes are denoted as negative training examples. It needs to consider how to extract the node and the
edge features.

\end{itemize}
\fi

\subsection{Feature Set}
 We use a set of node- and edge-level features for the learning models. The following set of
node-level features (denoted by $N$) are used. Each feature is normalized by the maximum value of the corresponding feature so that
the values range between 0 to 1.\\
$\bullet$ Normalized number of citations received by an author\\
$\bullet$ Normalized h-index of an author\\
$\bullet$ Normalized number of coauthors of an author \\
$\bullet$ Fraction of papers by an author in each of the 24 fields\\
$\bullet$ Normalized number of papers written by an author\\
$\bullet$ Fraction of papers published by an author in each of the
five decades (between 1960-2009)

Further, given an edge $e = (x,y)$ in the co-authorship network, we additionally use the following edge-level features (denoted by $E$).
Each
feature is appropriately normalized to a value between 0 and 1.\\
$\bullet$ Fraction of papers coauthored by $x$ and $y$ in each of the five decades\\ 
$\bullet$ Normalized number of common coauthors of $x$ and $y$\\
$\bullet$ Fraction of papers authored by $x$ in the major field of $y$ \\
$\bullet$ Fraction of papers authored by $y$ in the major field of $x$


We refer to the combined set of both node- and edge-level features by $NE$. We provide this set $NE$ of node and
edge attributes as an input to the learning model which then takes care of determining how to combine them with the network structure to
make predictions \cite{Backstrom}.
Note that if we take the dataset till $t$ for training the model, all the features mentioned above will be calculated based on the
statistics of each vertex till $t$ in order to avoid information leakage.

\if{0}
\subsection{Edge strength function}
Another important aspect in SRW model is the choice of the edge strength function $f_w(.)$ to combine the weight (obtained after training
the
model) for each feature  and the value of the corresponding feature. The edge strength must be non-negative and differentiable. 
Backstrom and Leskovec \cite{Backstrom} experimented with two functions -- exponential edge strength and logistic edge strength, and showed
that the choice of the edge strength function does not seem to make a significant impact on performance. However, there was slight evidence
from our experiments that the logistic function performs better. Therefore, in this experiment we consider the logistic edge strength
$a_{xy}$ of an edge $(x,y)$ as follows. We start by taking the inner product of the weight vector $w$ and the feature vector
$\psi_{xy}$ of an edge $(x,y)$. Then to transform this into the desired domain, we apply the logistic function:
\begin{equation}
 a_{xy}=(1+exp(\psi_{xy}.w))^{-1}
\end{equation}

In the above discussion, the weight vector $w$ is determined by SRW during the training phase.

\subsection{Competing Methods}
We design three competing models using the SRW framework as follows:
\begin{itemize}

\item {\bf SRW+NF:}This model uses only the node-level feature vector $NF$.
\item {\bf SRW+NF:}This model uses only the edge-level feature vector $EF$.
\item{\bf SRW+F:} This model uses only the feature vector $F$.
\item {\bf SRW+FB:} This model, besides $F$ also includes a binary feature $B$ that checks whether a pair of nodes $(x,y)$ belong to at
least one common
ego-centric circle or not.

\item{\bf SRW+FBC:} This mode besides $F$ and $B$ also includes a numeric feature $C$ indicating the number of common circles a pair of
nodes $(x,y)$ is a part of.
\end{itemize}

\fi

\begin{table*}[!ht]
\centering
\caption{Comparison of BIGCLAM (BIG), coordinate ascent method (CA) \cite{McAuley,Julian} and our model (CIRC) after including their
detected circle
information into the feature set of Linear Regression (LR) and Supervised
Random Walks (SRW) frameworks across three time periods and
different feature sets (N: node-level, E: edge-level, NE: node- and
edge-level, NEB: adding the binary circle information to NE, NEBC:
adding the numerical circle information to NEB).
}\label{table:auc}
\scalebox{0.68}{
\begin{tabular}{|c||c|l|l|l|l|l|l|l|l||l|l|l|l|l|l|l|l|l|}
\hline
\multirow{4}{*}{\begin{tabular}[c]{@{}c@{}}Time\\ period\end{tabular}} & \multicolumn{18}{c|}{Area Under the ROC Curve (AUC)}
\\\cline{2-19} 
     & \multicolumn{9}{c||}{Linear Regression (LR)}  & \multicolumn{9}{c|}{Supervised Random Walks (SRW)} \\ \cline{2-19} 
                                                                          & \multirow{2}{*}{N}    & \multicolumn{1}{c|}{\multirow{2}{*}{E}}
& \multicolumn{1}{c|}{\multirow{2}{*}{NE}} & \multicolumn{3}{c|}{NEB} & \multicolumn{3}{c||}{NEBC} & \multicolumn{1}{c|}{\multirow{2}{*}{N}}
& \multicolumn{1}{c|}{\multirow{2}{*}{E}} & \multicolumn{1}{c|}{\multirow{2}{*}{NE}} & \multicolumn{3}{c|}{NEB} & \multicolumn{3}{c|}{NEBC}
\\ \cline{5-10} \cline{14-19} 
                                                                          &                       & \multicolumn{1}{c|}{}                  
& \multicolumn{1}{c|}{}                    & BIG   &  CA      & CIRC        & BIG & CA        & CIRC         & \multicolumn{1}{c|}{}

& \multicolumn{1}{c|}{}                   & \multicolumn{1}{c|}{}                    & BIG   &  CA       & CIRC        & BIG  & CA       &
CIRC   
 
\\ \hline

1996-1999   &   0.5872 &   0.5914  &   0.6451  &   0.6569 & 0.6689 & 0.6791  &  0.6989 & 0.7195 &  {\bf 0.7235}  & 0.6332 & 0.6478 & 0.7659
& 0.7908 & 0.7895 & 0.8275 & 0.7971 &  0.8296 & {\bf 0.8303} \\\hline
2001-2004  & 0.5890  &   0.5907   &  0.6528  &   0.6529 & 0.6437 & 0.6659  &  0.6845 & 0.7011 & {\bf 0.7012}  &0.6419  &    0.6514  &   
0.7591 & 0.8067 & 0.8035 & 0.8249 &  0.8098 & 0.8149 & {\bf  0.8356}\\\hline
2006-2009   &   0.5916  &  0.5891  &   0.6436 &    0.6439 & 0.6510 & 0.6509  &  0.6905 & 0.7001 & {\bf 0.7198}    &  0.6360  &    0.6608  & 
 0.7609 & 0.8001 & 0.8101 & 0.8295 & 0.8111 & 0.8279 &  {\bf 0.8321}\\\hline 
Average & 0.5893 & 0.5904 & 0.6472 & 0.6512 & 0.6545 & 0.6653 & 0.6913 & 0.7069 & {\bf
  0.7148} & 0.6370 & 0.6533 & 0.7620 & 0.7992 & 0.8101 & 0.8273 & 0.8060 & 0.8279 & {\bf 0.8327}\\\hline
\end{tabular}}

\scalebox{0.68}{
\begin{tabular}{|c||c|l|l|l|l|l|l|l|l||l|l|l|l|l|l|l|l|l|}
\hline
\multirow{4}{*}{\begin{tabular}[c]{@{}c@{}}Time\\ period\end{tabular}} & \multicolumn{18}{c|}{$Prec@20$}            
                                                                                                                                            
                                                                                                                                          \\
\cline{2-19} 
                                                                          & \multicolumn{9}{c||}{Linear Regression (LR)}

                                                                                                 & \multicolumn{9}{c|}{Supervised Random
Walks (SRW)}                                                                                                                                
   
 \\ \cline{2-19} 
& \multirow{2}{*}{N}    & \multicolumn{1}{c|}{\multirow{2}{*}{E}}
& \multicolumn{1}{c|}{\multirow{2}{*}{NE}} & \multicolumn{3}{c|}{NEB} & \multicolumn{3}{c||}{NEBC} & \multicolumn{1}{c|}{\multirow{2}{*}{N}}
& \multicolumn{1}{c|}{\multirow{2}{*}{E}} & \multicolumn{1}{c|}{\multirow{2}{*}{NE}} & \multicolumn{3}{c|}{NEB} & \multicolumn{3}{c|}{NEBC}
\\ \cline{5-10} \cline{14-19} 
                                                                          &                       & \multicolumn{1}{c|}{}                  
& \multicolumn{1}{c|}{}                    & BIG   &  CA      & CIRC        & BIG & CA        & CIRC         & \multicolumn{1}{c|}{}

& \multicolumn{1}{c|}{}                   & \multicolumn{1}{c|}{}                    & BIG   &  CA       & CIRC        & BIG  & CA       &
CIRC   
 
\\ \hline

1996-1999 & 0.137 & 0.124 & 0.152 &    0.155 & 0.161 & 0.158 & 0.164& 0.173& {\bf 0.177} &  0.165 &  0.172 & 0.201 & 0.205 & 0.209 & 0.210 &
0.207 & 0.215 & {\bf 0.223} \\\hline
2001-2004 & 0.141 & 0.143 & 0.156 &    0.162 & 0.159 & 0.169 & 0.175 & 0.175 & {\bf 0.185} &  0.158 &  0.163 & 0.198 & 0.200 & 0.210 & 0.209
& 0.215 & 0.220 & {\bf 0.225}\\\hline
2006-2009 & 0.147 & 0.142 & 0.161 &    0.162 & 0.165 &  0.171 & 0.179 & 0.178 &{\bf 0.189} &  0.161 &  0.169 & 0.199 & 0.208& 0.209  &0.212
& 0.211& 0.217 &{\bf 0.224}\\\hline
Average & 0.142 & 0.136 & 0.156 & 0.160 & 0.162 & 0.166 & 0.173 & 0.175 & {\bf 0.184}
& 0.161 & 0.168 & 0.199 & 0.204 & 0.209 & 0.210 & 0.211 & 0.217& {\bf 0.224}\\\hline
\end{tabular}}
\vspace{-0.2cm}
\end{table*}

\subsection{Evaluation Methodology}\label{subsec:evaluation}
In order to demonstrate that predictions are robust irrespective of the time stamp considered for dividing the
dataset into training and test sets, we run the competing models in three different time periods: (i) the dataset till 1995 is considered
for training and the accuracies
of the models are measured by comparing the new edges formed between 1996-1999, (ii) similarly, the dataset till 2000 for training and
2001-2004 for checking the accuracy, and (iii) the dataset till 2005 for training and 2006-2009 for checking the accuracy. 

In each time stamp, we evaluate the methods on the test set, considering two performance metrics: 
the Area under the ROC curve ($AUC$) and the Precision at Top 20 ($Prec@20$), i.e., for each node $s$, what fraction of top 20 nodes
suggested by each model actually receive links from $s$ later. 
This measure is particularly appropriate in the context of link-recommendation where we present 
a user with a set of suggested coauthors and aim that most of them are correct.

\subsection{Performance Evaluation}
We compare the predictive performance of two learning models including the circle information in three different time periods as
mentioned in Section \ref{subsec:evaluation}. We iterate each of these collaboration prediction models using different sets of features:
(i) only node-level features ({\em Model: N}), (ii) only edge-level features ({\em Model: E}), (iii) both node and edge level features ({\em
Model: NE}), (iv) besides node and edge level features, including a binary feature $B$ that checks whether a pair of nodes
$(x,y)$ belong to at least one common ego-centric circle or not ({\em Model: NEB}), and (v) besides node-level and edge-level features and
the binary circle information, including a numeric feature $C$ indicating the number of common circles a pair of nodes $(x,y)$
is a part of ({\em Model: NEBC}). The circles are detected by our model, the coordinate ascent method (CA) \cite{McAuley,Julian} and BIGCLAM
separately.

Table \ref{table:auc} shows the performance of these two prediction models with different feature sets. We notice that edge features are
more effective than node features, and the performance improves incrementally after combining different
features together. A general observation is that
inclusion of circle information in the feature set improves the
performance of both the prediction models irrespective of the time
periods. For instance, it improves the performance by $9.87\%$ and $15.25\%$ on average in
terms of $AUC$ and $Prec@20$ respectively compared to the
case, where the circle information is not present ($NE$).

We further observe that the inclusion of circle information detected by our model significantly outperforms the case where the circles are 
obtained by BIGCLAM and CA in
each time stamp. Including the binary circle information ($NEB$) from our model achieves an average AUC improvement of 2.16\%  and
3.51\%  respectively for LR and SRW models (similarly, in terms of  $Prec@20$, the improvement is 3.75\%  and 2.94\% respectively for LR and
SRW models) compared to BIGCLAM (CA).  

Further, including the count of common circles for a node pair ($NEBC$) in the feature set leads both LR
and SRW to achieve even better performance. We observe an average AUC improvement of 3.41\% (1.11\%) and 3.31\% (0.57\%) respectively for LR
and SRW models
using our circle information as compared to that obtained from BIGCLAM (CA) (similarly, in terms of  $Prec@20$, the improvement is 6.35\%
(5.14\%) and
6.16\% (3.22\%) respectively for LR and SRW models). 

%
%

\vspace{-3mm}
\section{Conclusions and Future Work}\label{sec:conclusion}

Circles allow us to organize the overwhelming volumes of data generated by an author's personal academic network. In this work, we proposed
a simple yet effective method of detecting ego-centric circles in co-authorship networks. 
However, the proposed method is applicable to any general ego network given a suitable set of features. Our model is unsupervised
and combines node attributes and node similarities to identify circles that resemble communities in real networks. Experiments with four
state-of-the-art overlapping community detection algorithms showed that our method outperformed these baseline
algorithms. Further, a task based evaluation achieved a superior performance after inclusion of the circle information detected by our
model.

In future, we would like to develop a semi-supervised version of our algorithm that makes use of manually labeled data. 
Although most
authors may not want to label the circles manually, it would be highly desirable to make use of the information from those few who do.
Additionally, we would also like to apply the proposed method on the other datasets.

\vspace{-2mm}
\section{Acknowledgments}
The first author is financially supported by Google India PhD fellowship.

\vspace{-2mm}
\bibliographystyle{abbrv}

\end{document}